\documentstyle[11pt,newpasp,twoside,epsf]{article}
\def\edcomment#1{\iffalse\marginpar{\raggedright\sl#1\/}\else\relax\fi}
\reversemarginpar
\begin{document}
\title{On metal abundances in QSO absorbers}
\author{Sergei A. Levshakov}
\affil{Department of Theoretical Astrophysics, Ioffe Physico-Technical
Institute, Politechnicheskaya Str. 26, 194021 St. Petersburg, Russia}
\begin{abstract}
I review our recent results concerning the metal abundances
in quasar absorption-line spectra obtained with the 
UVES/VLT, HIRES/ KECK, and STIS/HST
spectrographs. The analysis is based on the Monte Carlo inversion
procedure aimed at the restoring the physical parameters
and the 1D distributions of the velocity and density fields
in quasar absorbers. 
A functional dependence of the line-of-sight velocity dispersion,
$\sigma_{\rm v}$, on the absorber linear size, $L$, was found: the
majority of the analyzed systems follow the scaling relation
$\sigma_{\rm v} \sim (N_{\rm H}\,L)^{0.3}$ (with $N_{\rm H}$ being
the total gas column density). This may indicate that the metal absorbers
are associated with virialized systems like galaxies or their halos.
A metal content enhancement (up to $Z \sim 4 Z_\odot$) 
was observed in small-size absorbers ($L \la 0.4$ kpc), 
whereas decreasing metal abundances were found in systems
with increasing $L$.
The lowest metallicity ([C/H] $\simeq -3.0$) 
was detected in the $z_{\rm abs} = 2.917$ Lymna-limit system (LLS) 
with $N$(\ion{H}{i}) = $3.2\times10^{17}$ cm$^{-2}$ and $L \simeq 140$ kpc. 
The relative abundance ratio of [Si/C] $\simeq 0.35$,
measured in this LLS, seems to indicate that
the initial mass function for the stellar population,
produced the observed metallicity pattern, was
constrained to intermediate masses, $M_{\rm up} \la 25M_\odot$.
\end{abstract}

\section{Introduction}

Metal absorption-line systems in quasar spectra 
are traditionally used to study different physical processes
in the intergalactic matter (IGM) over a wide range of redshifts,
back to the time when the Universe was less than 7\%
of its present age.
It is supposed that at $z \ga  10$ the IGM
was pre-enriched by the first Population III (Pop III)
stars formed from gas with zero
metallicity (e.g., Bromm et al. 2001; Nakamura \& Umemura 2001)
and enriched later due to
disruption of low-mass protogalaxies (e.g., Madau et al. 2001)
at $6 \la z \la 10$, and due to ejection of metals
from massive galaxies at $z \la 6$ (e.g., Aguirre et al. 2001;
Scannapieco et al. 2002).

The simulations of collapse and fragmentation of
primordial gas clouds suggest that the first generation
stars may have been
very massive, $M_\ast \ga 100M_\odot$ (Abel et al. 2000;
Bromm et al. 2001). However, the calculations by Nakamura \& Umemura (2001)
produced a bimodal initial mass function (IMF) for
Pop~III stars with a second peak   
at $1-2M_\odot$. According to these calculations the first generation
supernova ($10M_\odot < M_\ast < 35M_\odot$) and pair-instability supernova
($140M_\odot < M_\ast < 260M_\odot$) can produce metallicity in the range
$10^{-4}$ to $10^{-3}Z_\odot$.
Since the element yields and production
rate differ significantly for the massive ($10-35M_\odot$)
and very massive
($140-260M_\odot$) stars (Woosley \& Weaver 1995; Umeda \& Nomoto 2001;
Heger \& Woosley 2002), the Pop~III IMF can be
constrained by measuring the
abundance ratios in low metallicity cosmic objects.

Unfortunately, the computational
methods usually applied to high resolution spectra of QSOs
lie quite often  behind the quality of observational
data and fail to extract from them all encoded information.
The common processing method consists of the deconvolution
of complex absorption
profiles into an arbitrary number of separate
components (assuming a constant gas density within each of them),
which are then fitted to Voigt profiles.
However, in many cases this procedure may not correspond to
real physical conditions: observed complexity
and non-similarity of the profile shapes of different ions
indicate that these systems are in general absorbers
with highly fluctuating density and velocity fields
tightly correlated with each other.
The need for more sophisticated procedures of data analysis is
therefore obvious.

In recent years, it has been shown that accounting for the correlations
in the velocity field ({\it mesoturbulence}) may change the interpretation
of the absorption spectra substantially (e.g., Levshakov et al. 2000a;
Levshakov et al. 2002, hereafter LACM).
Our first inversion codes, ---  the Reverse Monte Carlo
(Levshakov et al. 1999a), and the Entropy-Regularized Minimization
(Levshakov et al. 1999b), --- have been developed for a model of
a stochastic velocity field neglecting any density fluctuations. They
have been applied to the analysis of the \ion{H}{i}
and \ion{D}{i} lines and/or to the metal absorption lines with
similar profiles when different species trace the same volume elements
independently of the density fluctuations.
Later on, we extended this study to the inverse
problem for a model of compressible turbulence when one observes
non-similar profiles of different atoms and/or ions
from the same absorption-line system. The developed procedure, called
the Monte Carlo inversion (MCI), is described in Levshakov et al.
(2000b, hereafter LAK), and its advanced version in LACM.

The MCI code has been applied to the analysis of 
metal absorbers with
$4\times10^{13}$ cm$^{-2}$ $< N$(\ion{H}{i}) $< 5\times10^{17}$ cm$^{-2}$
from the following
QSO spectra: J2233--606 (LACM), Q0347--3819 and
APM BR J0307--4945 (Levshakov et al. 2003a), HE 0940--1050
(Levshakov et al. 2003b), HE 0515--4414 (Agafonova et al. 2003a),
and PKS 0528--250 (Agafonova et al. 2003b). 
A brief description of the obtained results is given below.

\section{Model assumptions and the MCI procedure}

Numerous hydrodynamical calculations performed in the
previous decade have shown that the QSO absorption lines arise more likely
in the smoothly fluctuating intergalactic medium
in a network of sheets, filaments, and
halos (e.g., Cen et al. 1994; Miralda-Escud\'e et al. 1996;
Theuns et al. 1998).
A very important characteristic of the continuous absorbing medium is that
the contribution to any point within the line profile comes
not only from a single separate area (a `cloud')
but from all volume elements (`clouds')
distributed along the sightline within the 
absorbing region and having the same radial velocity
(see, for details, \S~2.2 in LAK).

The MCI procedure is based on the assumption that
all lines observed in a metal system arise
in a continuous absorbing gas slab of a thickness $L$ with
a fluctuating gas density and a random velocity field.
We also assume that
within the absorber the metal abundances are constant,
the gas is optically thin for the ionizing UV radiation, and
the gas is in the thermal and ionization equilibrium.
The last assumption means that
the fractional ionizations of different ions are determined
exclusively by the gas density  and vary from point to point
along the sightline.
These fractional ionization variations are just the cause
of the observed diversity of profile shapes between ions of low- and
high-ionization stages.

It is well known that the measured metallicities depend in a crucial way
on the adopted background ionizing spectrum. We started in all cases with
the Haardt-Madau (HM) background ionizing spectra (Haardt \& Madau 1996)
computing the fractional ionizations and the kinetic temperatures
with the photoionization code CLOUDY (Ferland 1997).
If the fitting with the HM spectrum was impossible, 
we used another spectra.

The MCI procedure itself is implemented in the following way.
Within the absorbing region the radial velocity  $v(x)$ and
the total hydrogen density $n_{\rm H}(x)$
along the line of sight are considered as two random fields
which are represented by their sampled values
at equally spaced intervals $\Delta x$,
i.e. by the vectors
$\{ v_1, \ldots, v_k \}$ and $\{ n_1, \ldots, n_k \}$
with $k$ large enough ($\sim 150-200$)
to describe the narrowest components of the complex spectral lines.
The radial velocity is assumed to be normally distributed with the
dispersion $\sigma_{\rm v}$, whereas the gas density is distributed
log-normally with the mean $n_0$
and the second central dimensionless moment $\sigma_{\rm y}$
($y = n_{\rm H}/n_0$).
Both stochastic fields are calculated using 
the Markovian processes (see LAK for mathematical basics).
The set of parameters estimated
in the least-squares minimization  of the objective function
(see eqs.[29] and [30] in LAK) includes
$\sigma_{\rm v}$ and $\sigma_{\rm y}$ along with
the total hydrogen column density $N_{\rm H}$,
the mean ionization parameter $U_0$,
and the metal abundances $Z_a$
for $a$ elements observed in a given absorption-line system.

The computations are carried out in two steps: firstly a point
in the parameter space $\{N_{\rm H}, U_0, \sigma_{\rm v}, \sigma_{\rm y},
Z_{\rm a}\}$  is chosen at random and then an optimal configuration
of $\{v_i\}$ and $\{n_i\}$ for this parameter set is searched for.
These steps are repeated till a minimum of the objective function
($\chi^2 \sim 1$ per degree of freedom) is achieved.
To optimize the configurations of  $\{v_i\}$ and $\{n_i\}$,
the simulated annealing algorithm with Tsallis acceptance
rule (Xiang et al., 1997) and an adaptive annealing temperature choice
is used (details are given in LACM).

\begin{figure}
\vspace{-0.45cm}
\plottwo{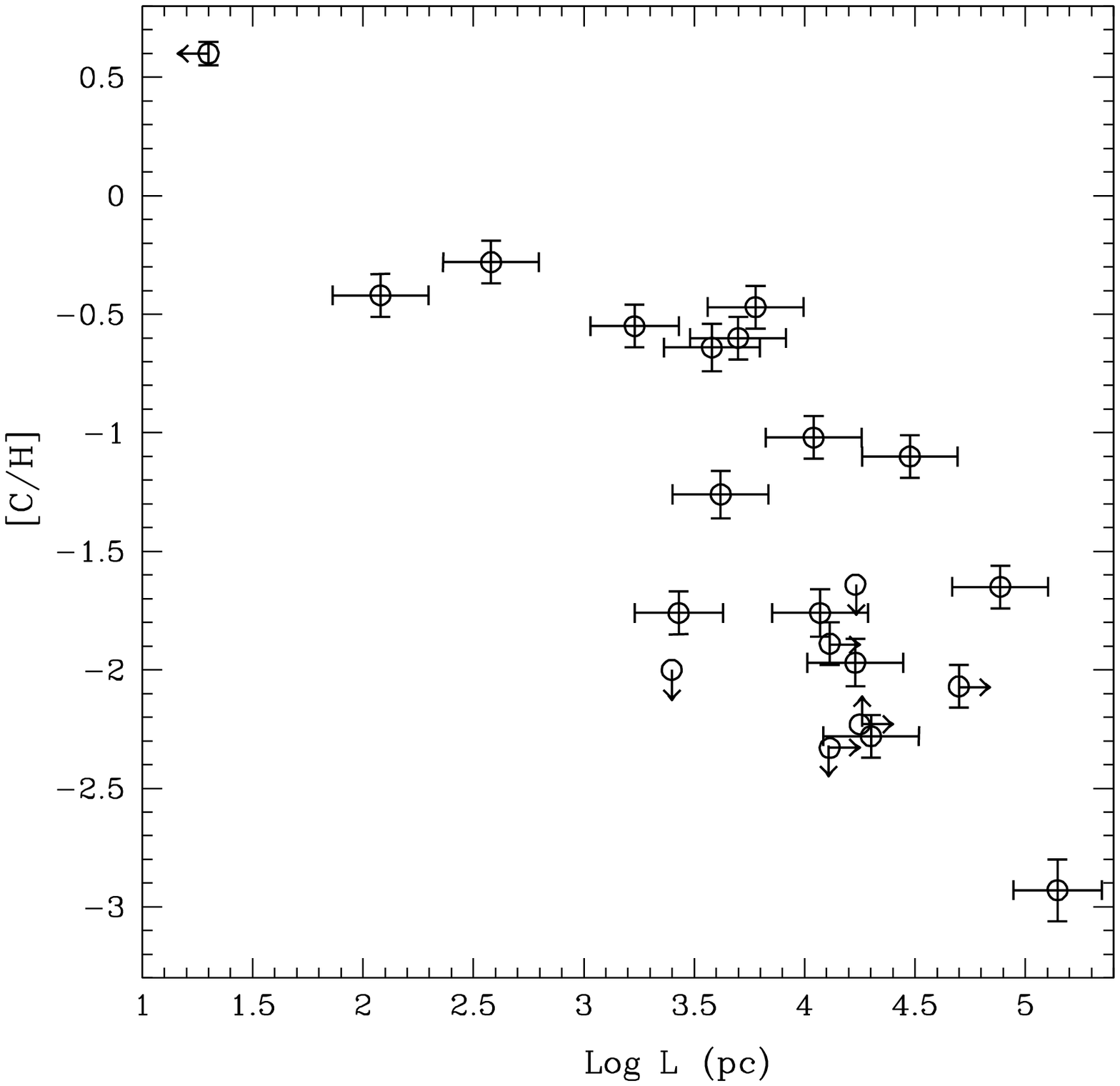}{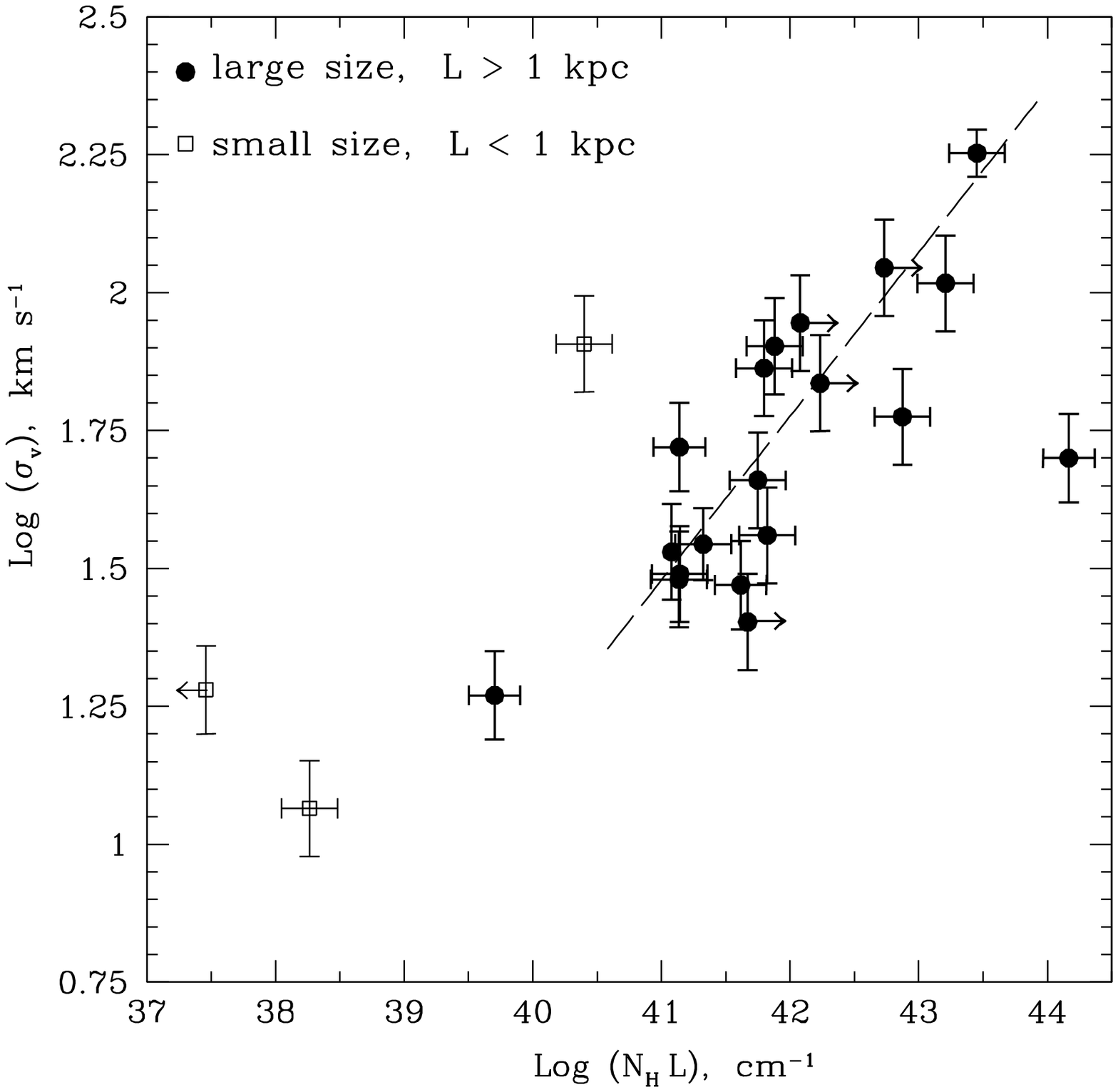}
\vspace{-0.5cm}
\caption{({\it left panel}) Carbon abundances  
plotted against the logarithmic linear
size of the corresponding systems as measured by the
MCI procedure. Small size QSO absorbers seem to have systematically 
higher metal content than systems with larger sizes.\\
({\it right panel}) Plot of the line of sight velocity 
dispersion $\log (\sigma_{\rm v})$ vs.  $\log (N_{\rm H}\,L)$ for
the same sample of the metal systems.
The dashed line is the linear regression $\log (\sigma_{\rm v}) \propto
\kappa\,\log (N_{\rm H}\,L)$ calculated for the points shown 
as filled circles having both horizontal error bars.
The slope $\kappa$ is equal to $0.30\pm0.03$ (1 $\sigma$ c.l.).
Open squares represent HVC-like systems.
}
\end{figure}

\section{Main results}

From the above mentioned QSO spectra, we have selected 22 absorption
systems suitable for the inversion analysis. 
Our results show
that the QSO intervening systems are a {\it heterogeneous} population
which is formed by at least three groups of absorbers~:
(1) extended metal-poor gas halos of distant galaxies;
(2) gas in dwarf galaxies; and
(3) metal-enriched gas arising from the inner
galactic regions and condensing into the clouds within 
the hot galactic halo (galactic fountain).
While the interpretation of a single system is sometimes subject to large
uncertainties, the existence of a wide spread of
properties in the different systems is certainly proved.

Figure~1 ({\it left panel}) shows an example of the measured carbon
abundances [C/H]\footnote{Using the customary definition
[X/H] = log\,(X/H) -- log\,(X/H)$_\odot$. Photospheric solar abundances
are taken from Grevesse \& Sauval (1998) and Holweger (2001).}
versus logarithmic sizes of the studied systems.
Systematically higher metal abundances are seen in compact systems
with linear sizes $L \la 0.4$ kpc. The highest metallicity of
[C/H] $= 0.59\pm0.08$ was measured in the associated system at
$z_{\rm abs} = 1.697$ toward HE 0515--4144, the linear size of this
system is only about 1 pc (Agafonova et al. 2003a). 
The lowest carbon content ([C/H] $= -2.93\pm0.13$)
was found in the LLS at $z_{\rm abs} = 2.917$ toward HE 0940--1050,
which shows $L \simeq 140$ kpc (Levshakov et al. 2003b).
In this LLS we also measured [Si/C] = $0.35\pm0.15$ and
[Al/Si] = $-0.18\pm0.13$. These values certainly rule out the
enrichment by very massive stars explosions and constrain the
stellar masses by $\sim 25 M_\odot$.
If, nevertheless, the Pop III stars were massive ($M > 140M_\odot$),
then the average metagalactic
metallicity produced by their explosions should be much
lower than $10^{-3}Z_\odot$.

If metal systems are formed in gas clouds gravitationally bound with
intervening galaxies, the internal kinematics of the QSO absorbers should be
closely related to the total masses of the host galaxies. In case of galactic
population, different types of galaxies show different scaling relations
between the linear size
and the velocity width of emission lines (e.g., Mall\'en-Ornelas
et al. 1999). Possible correlation between
the absorber linear size $L$ and its
line-of-sight velocity dispersion $\sigma_{\rm v}$
was also mentioned in LACM.

The correlation between $\sigma_{\rm v}$ and $L$ stems from the
virial theorem which states~:  
$\sigma^2_{\rm v} \sim {M}/{L} \sim n_0\,L^2 = N_{\rm H}\,L$.
Assuming that the gas systems are in quasi-equilibrium,
one can expect $\sigma_{\rm v} \sim (N_{\rm H}\,L)^{1/2}$.

In Figure~1 ({\it right panel}) we examine our systems by comparing 
their kinematics ($\sigma_{\rm v}$)
with measured sizes ($L$) and total gas column densities ($N_{\rm H}$).
It is seen that
in the $\log (\sigma_{\rm v})$ versus $\log (N_{\rm H}\,L)$ diagram, most
systems with linear sizes $L > 1$ kpc lie along the line with the slope
$\kappa = 0.30\pm0.03$ (1 $\sigma$ c.l.).

Taking into account that we know neither the impact parameters nor the halo
density distributions, this result can be considered as a quite good fit to
the expected relation for the virialized systems. 
Hence we may conclude that most absorbers with $L > 1$ kpc are
gravitationally bound with systems that appear to be in virial
equilibrium at the cosmic time when the corresponding Ly$\alpha$ absorbers
were formed. 
If the most metal
absorbers identified in the QSO spectra arise in the galactic
systems then the question
to what degree
the intergalactic matter is metal enriched remains  open.

The absorption systems shown in Figure~1 ({\it right panel}) by open squares
exhibit characteristics very similar to that observed   
for different types of HVCs in the Milky Way
and may be interpreted as the high-redshift counterparts
of these Galactic objects.
The most clear example of such high-$z$ HVC is the 
mentioned above very metal-rich system
at $z_{\rm abs} = 1.697$
with the neutral hydrogen column density
$N$(\ion{H}{i}) $= 4.4\times10^{13}$ cm$^{-2}$. The system
shows absorption from
highly ionized transitions of \ion{C}{iii}, \ion{C}{iv},
\ion{N}{v}, \ion{O}{vi}, \ion{Si}{iv}, and probably \ion{S}{vi}.
We found that only a power law ionizing spectrum
($J_\nu \propto \nu^{-1.5}$)
is consistent with the observed sample of the line profiles,
i.e., the system was probably blown out from the QSO/host galaxy
by a cumulative effect of the supernova
in the starburst and the quasar activity. 
Such metal-enriched outflowing gas is supposed to
give rise to some HVCs observed in the
Milky Way halo (Bregman 1980; Wakker 2001).

Some of the small-size metal systems show very strong metallicity
gradients. A system at $z_{\rm abs} = 1.385$ toward HE 0515--4414
is a good example. Here we measured 
$N$(\ion{H}{i}) $\simeq 5.3\times10^{13}$ cm$^{-2}$,
[C/H] $\simeq -0.5$, and $L \simeq 2$ kpc. 
This system is probably embedded in an extremely metal-poor
halo with $N$(\ion{H}{i}) $\simeq 1.1\times10^{15}$ cm$^{-2}$,
[C/H] $< -4$, and $L \simeq 90$ kpc (Agafonova et al. 2003a).
The velocity shift between these subsystems is only
$\Delta v = 140$ km~s$^{-1}$. The velocity excess of the Galactic HVCs
is usually greater than 90 km~s$^{-1}$, which is consistent with this
$\Delta v$ value. 

In conclusion, we would like to emphasize that 
in spite of a utterly inhomogeneity in metal abundances
revealed in the Ly$\alpha$ absorbers, all of them are
perfectly described within pure photoionization models.
Our results show that the fraction of the supposed shock-heated
hot gas with temperature $T > 10^5$ K is negligible in the
analyzed absorbers, and that it is impossible to relate
unambiguously highly ionized absorption systems to the
`warm-hot' gas predicted in cosmological models (e.g.,
Cen \& Ostriker 1999; Dav\'e et al. 2001).
Each highly ionized system requires a comprehensive study in order to
determine its nature and origin.


\begin{references}
%
\reference Abel, T. G., Bryan, G. L., \& Norman, M. L. 2000, ApJ, 540, 39
%
\reference Agafonova, I. I., Baade, R., Levshakov S. A., \& Reimers, D.
2003a, in prep.
%
\reference Agafonova, I. I., Centuri\'on, M., Levshakov S. A., \& 
Molaro, P. 2003b, in prep.
%
\reference Aguirre, A., et al. 2001, 
ApJ, 560, 599
%
\reference Bregman, J. N. 1980, ApJ, 236, 577
%
\reference Bromm, V., Ferrara, A., Coppi, P. S., \& Larson, R. B. 2001,
MNRAS, 328, 969
%
\reference Cen, R., Miralda-Escude\'e, J., Ostriker, J. P., \&
Rauch, M. 1994, ApJ, 437, L9
%
\reference Cen, R., \& Ostriker, J. P. 1999, ApJ, 514, 1
%
\reference Dav\'e, R., et al. 2001, ApJ, 552, 473
%
\reference Ferland, G. J. 1997, A Brief Introduction to
Cloudy (Internal Rep.,; Lexington: Univ. Kentucky)
%
\reference Grevesse, N., \& Sauval, A. J. 1998, Space Sci. Rev., 85, 161
%
\reference Haardt, F., \& Madau, P. 1996, ApJ, 461, 20
%
\reference Heger, A., \& Woosley, S. E. 2002, ApJ, 567, 532
%
\reference Holweger, H. 2001, in Solar and Galactic Composition,
ed. R. F. Wimmer-Schweingruber, AIP Conf. Proceed.. 598, 23
%
\reference Levshakov, S. A., Kegel, W. H., \& Takahara, F. 1999a,
MNRAS, 302, 707 
%
\reference Levshakov, S. A., Takahara, F., \& Agafonova, I. I. 1999b,
ApJ, 517, 609
%
\reference Levshakov, S. A., Agafonova, I. I., \& Kegel, W. H.
2000a, A\&A, 355, L1
%
\reference Levshakov, S. A., Agafonova, I. I., \& Kegel, W. H.
2000b, A\&A, 360, 833 [LAK]
%
\reference Levshakov, S. A., Agafonova, I. I., Centuri\'on, M., \&
Mazets, I. E. 2002, A\&A, 383, 813 [LACM]
%
\reference Levshakov, S. A., Agafonova, I. I., D'Odorico, S., Wolfe, A. M., \&
Dessauges-Zavadsky, M. 2003a, ApJ, 582, in press (astro-ph/0209328)
%
\reference Levshakov, S. A., Agafonova, Centuri\'on, M., \&
Molaro, P. 2003b, A\&A, in press (astro-ph/0210619)
%
\reference Madau, P., Ferrara, A., \& Rees, M. J. 2001, ApJ, 555, 92
%
\reference Mall\'en-Ornelas, G., Lilly, S. J., Crampton, D., \& Schade, D. 
1999, ApJ, 518, L83
%
\reference Miralda-Escud\'e, J., Cen, R.,
Ostriker, J. P., \& Rauch, M. 1996, ApJ, 471, 582
%
\reference Nakamura, F., \& Umemura, M. 2001, ApJ, 548, 19
%
\reference Scannapieco, E., Ferrara, A., \& Madau, P. 2002, ApJ, 574, 590
%
\reference Theuns, T., Leonard, A., Efstathiou, G.,
Pearce, F. R., \& Thomas, P. A.  1998, MNRAS, 301, 478
%
\reference Umeda, H., \& Nomoto, K. 2001, ASP Conf. Ser., 222, 45
%
\reference Wakker, B. P. 2001, ApJS, 136, 463
%
\reference Woosley, S. E., \& Weaver, T. A. 1995, ApJS, 101, 181
%
\reference Xiang, Y., Syn, D. Y., Fan, W., \& Gong, X. G. 1997,
Phys. Lett. A, 233, 216

\end{references}
\end{document}